\pdfoutput=1

\documentclass[11pt,twoside,a4paper,cmspaper,final,collab]{cms-tdr}
\def\svnVersion{471369P}\def\cmsCernNoTag{CERN-EP-2018-004}\def\cmsCernDate{\today}\def\cmsMessage{Published in the Journal of High Energy Physics as \href{http://dx.doi.org/10.1007/JHEP07(2018)161}{\doi{10.1007/JHEP07(2018)161.}}}

\begin{document}\cmsNoteHeader{FSQ-15-005}

\hyphenation{had-ron-i-za-tion}
\hyphenation{cal-or-i-me-ter}
\hyphenation{de-vices}
\RCS$HeadURL: svn+ssh://svn.cern.ch/reps/tdr2/papers/FSQ-15-005/trunk/FSQ-15-005.tex $
\RCS$Id: FSQ-15-005.tex 470898 2018-08-04 11:31:05Z pvanmech $

\providecommand{\NA}{\ensuremath{\text{---}}}
\providecommand{\CL}{CL\xspace}
\newlength\cmsTabSkip\setlength{\cmsTabSkip}{1ex}

\cmsNoteHeader{FSQ-15-005}
\title{Measurement of the inelastic proton-proton cross section at $\sqrt{s}=13\TeV$}

\date{\today}

\abstract{
A measurement of the inelastic proton-proton cross section with the CMS detector at a center-of-mass energy of $\sqrt{s} = 13\TeV$ is presented. The analysis is based on events with energy deposits in the forward calorimeters, which cover pseudorapidities of $-6.6 < \eta < -3.0$ and $+3.0 < \eta < +5.2$. An inelastic cross section of $68.6 \pm 0.5 \syst \pm 1.6\lum\unit{mb}$ is obtained for events with  $M_\mathrm{X} > 4.1\GeV$ and/or $M_\mathrm{Y} > 13\GeV$, where $M_\mathrm{X}$ and $M_\mathrm{Y}$ are the masses of the diffractive dissociation systems at negative and positive pseudorapidities, respectively.
The results are compared with those from other experiments as well as to predictions from high-energy hadron-hadron interaction models.
}

\hypersetup{%
pdfauthor={CMS Collaboration},%
pdftitle={Measurement of the inelastic proton-proton cross section at sqrt(s) =13 TeV},%
pdfsubject={CMS},%
pdfkeywords={CMS, physics, soft QCD}}

\maketitle

\section{Introduction}

Inclusive hadron-hadron cross sections are fundamental observables in high-energy particle, nuclear, and cosmic ray physics, and have been measured in experiments covering many orders of magnitude in center-of-mass energy, $\sqrt{s}$.  The hadron-hadron cross section can be decomposed into elastic and inelastic contributions (including diffractive and nondiffractive topologies), each of which can be described by nonperturbative phenomenological models, based on general principles such as unitarity and analyticity \cite{Barone:2002cv}.  These models have, however, large uncertainties when extra\-polating existing data to higher center-of-mass energies \cite{Donnachie:1992ny,Schuler:1993td,Schuler:1993wr}, and need to be constrained by measurements.

Precise measurements of the proton-proton ($\Pp\Pp$) cross sections at CERN LHC energies provide valuable input for phenomenological hadronic interaction models, as well as for the tuning of Monte Carlo (MC) event generators used in collider \cite{Sjostrand:2006za,Sjostrand:2007gs,Werner:2005jf,Ostapchenko:2004ss,Ostapchenko:2010vb} and cosmic ray physics \cite{Ulrich:2010rg}.  The value of the inelastic $\Pp\Pp$ cross section, $\sigma_{\text{inel}}$, is also used in the estimation of the number of simultaneous $\Pp\Pp$ interactions occurring in the same beam-bunch crossing (henceforward referred to as ``pileup'') and is thus an important quantity to properly control the backgrounds in $\Pp\Pp$ collision events measured at the LHC. It is also a crucial ingredient of Glauber models used to describe high-energy nuclear collisions \cite{Loizides:2017ack}.

At $\sqrt{s} = 7\TeV$, there have been several $\Pp\Pp$ cross section measurements at the LHC. The CMS Collaboration has measured an inelastic $\Pp\Pp$ cross section of $ 60.2 \pm 0.2 \stat \pm 1.1 \syst \pm 2.4\lum\unit{mb}$ for $\xi = M^2/s > 5 \times 10^{-6}$ (corresponding to $M > 15.7\GeV$) \cite{Chatrchyan:2012nj}. The phase space region for which this measurement was obtained excludes elastic and diffractive proton dissociation events with very low invariant mass of the dissociation system, $M$ being the invariant mass of the larger of the two diffractive dissociation systems.   The ATLAS~\cite{Aad:2011eu} and ALICE~\cite{Abelev:2012sea} Collaborations measured the inelastic $\Pp\Pp$ cross section at $\sqrt{s} = 7\TeV$ in the same phase space and reported consistent values. Several collaborations have extrapolated their measurements to the full inelastic phase space.  The values reported for the total inelastic $\Pp\Pp$ cross section at $\sqrt{s} = 7\TeV$ range from $66.9 \pm 5.3$ to $73.7 \pm 3.4\unit{mb}$\cite{Aad:2011eu,Aad:2014dca,Antchev:2011vs,Antchev:2013gaa,Antchev:2013iaa,Abelev:2012sea,Aaij:2014vfa}.  The TOTEM Collaboration, in particular, derived the inelastic cross section by exploiting the optical theorem, thereby avoiding model-dependent extrapolations for the low-mass diffraction contribution.  A value of $73.2 \pm 1.3\unit{mb}$ was obtained \cite{Antchev:2013iaa}.

At $\sqrt{s} = 8\TeV$, the TOTEM Collaboration also measured the total, elastic, and inelastic $\Pp\Pp$ cross sections~\cite{Antchev:2013paa}, reporting a value of $\sigma_{\text{inel}} = 74.7 \pm 1.7 \unit{mb}$.

At $\sqrt{s} = 13\TeV$, the ATLAS Collaboration reported a measurement of the inelastic $\Pp\Pp$ cross section of $68.1 \pm 0.6 \syst \pm 1.3\lum \unit{mb}$ for $\xi > 10^{-6}$ (corresponding to $M > 13\GeV$)~\cite{Aaboud:2016mmw}.  This value has been extrapolated to the total inelastic phase space, yielding $\sigma_{\text{inel}} = 78.1 \pm 0.6 \syst \pm 1.3\lum \pm 2.6\, \text{(extr)} \unit{mb}$, with the last number being the extrapolation uncertainty. Finally, the TOTEM Collaboration obtained a value for the inelastic cross section at $\sqrt{s} = 13\TeV$ of $\sigma_{\text{inel}} = 79.5 \pm 1.8 \syst \unit{mb}$ \cite{Antchev:2017dia}.

This paper presents a new measurement of the inelastic cross section in $\Pp\Pp$ collisions at $\sqrt{s} = 13\TeV$.  Data collected with the CMS forward calorimeters HF and CASTOR, described in Section \ref{sec:detector} and covering pseudorapidities $-6.6 < \eta < -3.0$ and $+3.0 < \eta < +5.2$, are analyzed.     These detectors provide sensitivity to a large part of the total inelastic cross section, including diffractive events with dissociated protons that produce particles only at forward rapidity, with the exception of low-mass diffraction and  events that happen to have rapidity gaps in the regions covered, including central exclusive production, \eg, diffractive dissociation mediated via double pomeron exchange.  The fiducial cross section is therefore measured in a phase space region excluding fractional momentum losses of the scattered protons $\xi_{\mathrm{X}} = M_{\mathrm{X}}^2/s < 10^{-7}$ and $\xi_{\mathrm{Y}} = M_{\mathrm{Y}}^2/s < 10^{-6}$, corresponding to $M_{\mathrm{X}} < 4.1\GeV$ and $M_{\mathrm{Y}} < 13\GeV$, where $M_{\mathrm{X}}$ and $M_{\mathrm{Y}}$ are defined as the invariant masses of the dissociated proton systems with negative and positive pseudorapidities, respectively.  The use of the CASTOR forward calorimeter allows the extension of this type of measurement to a low mass region so far unexplored.

\section{The CMS detector}
\label{sec:detector}

CMS uses a right-handed coordinate system, with the origin at the nominal interaction point, the $x$ axis pointing to the center of the LHC, the $y$ axis pointing up (perpendicular to the LHC plane), and the $z$ axis along the anticlockwise-beam direction.

The central feature of the CMS apparatus~\cite{Chatrchyan:2008zzk} is a superconducting solenoid of $6\unit{m}$ internal diameter, providing a magnetic field of $B = 3.8\unit{T}$ parallel to the beams. Within the solenoid volume are a silicon pixel and strip tracker (covering $\abs{\eta} < 2.5$), a lead tungstate crystal electromagnetic calorimeter ($\abs{\eta} < 3$) , and a brass and scintillator hadron calorimeter ($\abs{\eta} < 3$), each composed of a barrel and two endcap sections.

Extensive forward calorimetry complements the coverage provided by the barrel and endcap detectors.  The forward hadron (HF) calorimeters are located on each side of the detector, covering the range $3.0 < \abs{\eta} < 5.2$, and are each composed of 18 steel azimuthal ($\phi$) wedges, with embedded quartz fibers running parallel to the beam direction. Each wedge is subdivided into 13 segments in $\eta$, called towers.  The very forward angles are covered at one end of CMS ($-6.6 < \eta < -5.2$) by the CASTOR calorimeter \cite{Andreev:2010zzb}. This detector, consisting of tungsten absorbers and quartz detection planes, is segmented in 16 $\phi$-sectors and into 2 electromagnetic and 12 hadronic modules along the beam line, corresponding to a total of 224 cells. For operational reasons the CASTOR calorimeter was only partially included in the detector setup during the run periods considered in this analysis, and the data with CASTOR included were taken with the solenoid switched off ($B = 0\unit{T}$).

A more detailed description of the CMS detector can be found in Ref.~\cite{Chatrchyan:2008zzk}.

\section{Physics models}
\label{sec:MCmodels}

Various Monte Carlo event generators are used to correct the measured cross section for acceptance and instrumental effects, as well as to compare the final results to different hadron-hadron interaction model predictions.

The \PYTHIA MC generator~\cite{Sjostrand:2006za,Sjostrand:2007gs} uses the Donnachie--Landshoff (DL) parametrization~\cite{Donnachie:1992ny} for the total hadron-hadron cross section, and provides different approaches to determine the elastic and diffractive contributions (unless stated otherwise, the default Schuler--Sj\"ostrand (SS) model~\cite{Schuler:1993td,Schuler:1993wr} is used). The difference between the total cross section and the sum of the elastic and diffractive cross sections is used to normalize the nondiffractive part, which is generated through a regularization of perturbative multi-parton interaction cross sections. Event samples are generated with different underlying event tunes: the Z2*~\cite{Chatrchyan:2013gfi} tune is used  for the \PYTHIA 6 (version 6.426) sample, while \PYTHIA 8 (version 8.205) samples are generated with the Monash~\cite{Skands:2014pea} and CUETP8M1~\cite{Khachatryan:2015pea} tunes. These differ in the parameters used to describe initial- and final-state parton showers, multi-parton interactions, and hadronization. Samples are generated with the Monash DL (using the Donnachie--Landshoff diffraction model~\cite{Donnachie:1984xq}) and Minimum--Bias--Rockefeller (MBR)~\cite{Ciesielski:2012mc} tunes. The MBR approach is based on a phenomenological renormalized Regge theory~\cite{Goulianos:2004as,Goulianos:2002vm}.

Two models based on Regge--Gribov phenomenology~\cite{Gribov:1968fc,Gribov:1968uy}, including the Abramowski--Gribov--Kancheli cutting rules~\cite{Abramovsky:1973fm}, are also used: \textsc{Epos}~\cite{Werner:2005jf} (version 1.99) with its LHC tune and \textsc{QGSJet-II} (version 04)~\cite{Ostapchenko:2004ss,Ostapchenko:2010vb}.

\section{Event selection and reconstruction}
\label{sec:detectorlevelanalysis}

\subsection{Event selection}

This analysis is based on $\Pp\Pp$ collision data collected with the CMS detector in 2015 at $\sqrt{s} = 13\TeV$, during several running periods with low pileup. Runs for which the solenoid was off ($B = 0\unit{T}$), as well as runs with the solenoid at its nominal field strength ($B = 3.8\unit{T}$) are analyzed. The former runs, with CASTOR included in the detector setup, are used because of its larger acceptance for inelastic collisions.

The total integrated luminosity recorded in these runs amounts to 40.8 (28.0)\mubinv for $B=3.8\unit{T}$ ($B = 0\unit{T}$).  The measurement of the integrated luminosity for the $B=3.8\unit{T}$ data is based on the pixel tracker, and has been calibrated by means of a dedicated analysis of van der Meer scans~\cite{CMS-PAS-LUM-15-001} with an accuracy of 2.3\%.
The integrated luminosity for the $B=0\unit{T}$ sample is obtained by normalizing the fiducial cross section found at  $B = 3.8\unit{T}$ to the one at $B=0\unit{T}$ in the same phase space.

The CMS data acquisition was triggered \cite{Khachatryan:2016bia} by the presence of both beams in the interaction point (``zero bias''), signaled by beam pick-up monitors located at $\pm$175\unit{m} from the interaction point.  Additional triggers requiring the presence of only one beam (``single bunch'') or no beams (``empty bunch'')  were used to study beam-gas, electronic noise, and other backgrounds.  These triggers are 100\% efficient for the event selection under consideration.   A first sample of inelastic events is then selected offline by requiring an energy deposit above 5\GeV in either of the two HF calorimeters. This threshold was optimized by studying detector noise in events without beam. The rate of selected events from single bunches was found to be consistent with the one from empty bunches. This demonstrates that the presence of the beam on one side does not generate more background events compared to no beam, and that the total background contribution can be estimated from the empty bunch events alone.

The presence of the CASTOR calorimeter in the $B=0\unit{T}$ data sample allows a larger coverage of the phase space for inelastic $\Pp\Pp$ collisions.  In this case, inelastic events are selected offline by requiring either an energy deposit above 5\GeV in either of the two HF calorimeters, or an energy deposit above 5\GeV in CASTOR.

\subsection{Correction for noise and pileup}
\label{sec:datacorrection}

The selected number of inelastic events is first corrected for the contribution of detector noise.  The corrected number of interactions is obtained as
\begin{equation}
N_{\text{cor}} = N_{\mathrm{ZB}} [ (F_{\mathrm{ZB}} - F_{\mathrm{EB}}) + F_{\mathrm{EB}}  (F_{\mathrm{ZB}} - F_{\mathrm {EB}})],
\label{eq:noisesubtraction}
\end{equation}
where $N_{\mathrm{ZB}}$ is the number of events triggered by the zero bias trigger and $F_{\mathrm{ZB}}$ and $F_{\mathrm{EB}}$ are the fractions of events triggered by the zero bias and empty bunch triggers that are selected offline by requiring an energy deposit above threshold, respectively.  In Eq.~(\ref{eq:noisesubtraction}), the second term on the right-hand side is a first-order correction for genuine signal events (with occurrence approximated by $F_{\mathrm{ZB}} - F_{\mathrm{EB}}$) overlaid with noise (with occurrence $F_{\mathrm{EB}}$).  Corrections of higher order in $F_{\mathrm{EB}}$ are found to be negligible. Table~\ref{tab:corxsec} includes an overview of the noise-subtracted fraction of events ($N_{\text{cor}} / (N_{\mathrm{ZB}}\, F_{\mathrm{ZB}})$) found in the various runs.

\begin{table}
\centering
\topcaption{Overview of the noise-subtracted fraction of events ($N_{\text{cor}}/(N_{\mathrm{ZB}}\, F_{\mathrm{ZB}})$), average pileup ($\lambda$), and fiducial cross section for all runs used in this analysis. The uncertainties in the noise-subtracted fraction of events and pileup have a negligible contribution to the uncertainty in the fiducial cross section, which is quoted with its statistical uncertainty only.}
\label{tab:corxsec}
\begin{tabular}{lccc}
Runs & $N_{\text{cor}}/(N_{\mathrm{ZB}}\, F_{\mathrm{ZB}})$ & $\lambda$ & Fiducial \\
$B = 3.8\unit{T}$ & (\%) &   &  cross section (\unit{mb}) \\
 & & &  $\xi > 10^{-6}$ \\ \hline
254989 & 98.5 & 0.52 & $67.35 \pm 0.05$ \\
255019 & 99.9 & 0.54 & $67.66 \pm 0.04$ \\
255029 & 99.3 & 0.54 & $67.50 \pm 0.04$ \\
\multicolumn{4}{c}{\ } \\
Runs & $N_{\text{cor}}/(N_{\mathrm{ZB}}\, F_{\mathrm{ZB}})$ & $\lambda$  & Fiducial \\
$B = 0\unit{T}$ & (\%) &  &  cross section (\unit{mb}) \\
 & & & $\xi_{\mathrm{X}} > 10^{-7}$ or $\xi_{\mathrm{Y}} > 10^{-6}$ \\\hline
247324  & 98.5 & 0.05 & $68.88 \pm 0.49$ \\
247920  & 98.9 & 0.34 & $68.63 \pm 0.08$ \\
247934  & 98.8 & 0.32 & $68.63 \pm 0.09$ \\
\end{tabular}
\end{table}

The number of events is further corrected for the effect of pileup. The observed number of $\Pp\Pp$ collisions per bunch crossing, $n$, follows a Poisson distribution, $P(n,\lambda)$, with average value $\lambda$. As the probability to find an interaction in a crossing with filled bunches is given by $N_{\text{cor}}/N_{\mathrm{ZB}}$, the probability to have no interaction can be obtained as $P(0,\lambda) \equiv \exp(-\lambda) = 1 - N_{\text{cor}}/N_{\mathrm{ZB}}$, which allows $\lambda$ to be determined from the data. With this information it is possible to correct the inelastic event count using the pileup correction factor:
\begin{equation}
f_{\mathrm{PU}} = \frac{\sum_{n=0}^{\infty} n P(n,\lambda)}{\sum_{n=1}^\infty P(n,\lambda)} = \frac{\lambda}{1-P(0,\lambda)}.
\label{eq:pileupcorrection}
\end{equation}
The values of $\lambda$ in the studied runs range from 0.05 to 0.54 and are given in Table~\ref{tab:corxsec}.  The average pileup value for run 247324 is substantially lower than for the other runs and demonstrates the robustness of the pileup correction, even though the statistical error on the fiducial cross section for this run, as obtained below, is larger due to the smaller number of recorded events.  As the beam intensity may vary from one proton bunch to another, the actual pileup correction is applied bunch-by-bunch. The total reconstructed number of interactions, corrected for the contributions of noise and pileup, is then given by
\begin{equation}
N_{\text{int}} =  \sum_{\text{bunches}} N_{\text{cor}}^{\mathrm{b}} \,  f_{\mathrm{PU}}^{\mathrm{b}},
\end{equation}
with  the number of noise-corrected events $N_{\text{cor}}^{\mathrm{b}}$ (Eq.~(\ref{eq:noisesubtraction})) and pileup correction factors $f_{\mathrm{PU}}^{\mathrm{b}}$ (Eq.~(\ref{eq:pileupcorrection})) calculated for individual bunches.

\section{Extraction of the fiducial inelastic cross section}
\label{sec:visxsection}

Before comparison to theoretical predictions, the experimental results need to be corrected for various detector effects, including the event selection
efficiency and the resolution in the energy measurement. Corrected results are obtained by means of a simulation of the CMS detector
based on \GEANTfour~\cite{Agostinelli:2002hh, Allison:2006ve,Allison:2016lfl}.

The event selection criteria at the detector level are optimized in order to obtain a sample of inelastic events in the largest possible phase space. Low-mass diffractive dissociation events will, however, escape the detector. Central diffractive dissociation is found to contribute at the level of 0.1--0.2 \unit{mb}, both from predictions obtained with the \PYTHIA MBR model and from counting events in the data with activity in the central detectors but without any energy deposited in the forward calorimeters, and is neglected in the further analysis.  Adapting the particle-level phase space to the detector acceptance results in a smaller correction of the results, and thus also in a smaller model dependence of the correction factors.  A precise definition of the phase space at the level of generated stable particles, for which corrected results are presented, is obtained as follows.

The collection of stable final-state particles (with proper lifetime $c\tau > 1\unit{cm}$) is divided into two systems, X and Y, based on the mean rapidity of the two particles separated by the largest rapidity gap in the event.  All particles on the negative (positive) side of the largest gap are assigned to the system X (Y). The invariant masses, $M_{\mathrm{X}}$ and $M_{\mathrm{Y}}$, of each system are calculated from the four-momenta of the individual particles, and are used to obtain the squared ratios of the mass over the center-of-mass energy, $\xi_{\mathrm{X}}$ and $\xi_{\mathrm{Y}}$.
For convenience, $\xi$ can be defined by $\xi = \max(\xi_{\mathrm{X}},\xi_{\mathrm{Y}})$. These Lorentz-invariant variables are well defined for any type of events (diffractive and nondiffractive) and are related to the size of the largest rapidity gap \cite{Barone:2002cv}.

The fiducial phase space can then be quantified at the stable-particle level by appropriate limits on $\xi_{\mathrm{X}}$ and $\xi_{\mathrm{Y}}$. These acceptance limits are obtained from a dedicated study based on the hadron-hadron interaction models mentioned in Section~\ref{sec:MCmodels} using fully simulated events, and are chosen such that the  factors required to correct the data to stable-particle level are close to unity, thereby minimizing the model dependence of the correction. An inelastic event is selected at the stable-particle level if $\xi > 10^{-6}$ for the selection based on the HF calorimeter alone, and, because the CASTOR calorimeter allows a lower $\xi$ limit on one side, if $\xi_{\mathrm{X}} > 10^{-7}$ or $\xi_{\mathrm{Y}} > 10^{-6}$ when the HF and CASTOR calorimeters are combined.

The relationship between the stable-particle level phase space definition and the detector-level off\-line selection can be quantified through efficiency and contamination factors.  The efficiency, $\epsilon_\xi$, is defined as the fraction of selected stable-particle level events that fulfill the detector-level offline selection criteria, while the contamination, $b_\xi$, is defined as the fraction of detector-level offline selected events that are not part of the considered stable-particle level phase space. The efficiency and contamination factors are calculated as the average over MC models and are found to be equal to 97.6\% and 0.6\% (98.6\% and 0.6\%) for the HF (HF+CASTOR) fiducial region, respectively.

Finally, the fiducial cross section is calculated as
\begin{equation}
\sigma = \frac{N_{\text{int}} (1-b_\xi)}{\epsilon_\xi \,  {\cal L}},
\end{equation}
and is given in Table~\ref{tab:corxsec} for all runs used in the analysis. The integrated luminosity, ${\cal L}$, of the $B=0\unit{T}$ runs has been rescaled in order to yield the same cross section for $\xi > 10^{-6}$ as measured in the $B=3.8\unit{T}$ runs, thus exploiting the more accurate luminosity determination in the latter runs.

\section{Systematic uncertainties}
\label{sec:systematics}

The following sources of systematic bias are investigated and the corresponding uncertainties are evaluated.

\begin{itemize}
\item \textbf{Model dependence.}  The efficiency and contamination factors are obtained from MC simulation. Although the phase space domains are chosen so as to minimize extrapolations, there is a remaining model dependence due to the matching of stable-particle level and detector-level selections.  These factors are defined as fractions and are therefore not very sensitive to the absolute magnitude of the inelastic cross section itself; instead they are affected by the way in which a hadronic system of a particular invariant mass fragments into individual particles that deposit energy in the calorimeters.  The uncertainty is therefore taken as the standard deviation of the correction factors obtained from all the models discussed in Section~\ref{sec:MCmodels}, even if some models do not describe the measured cross sections well.  
\item \textbf{HF and CASTOR energy scale uncertainties.}   The energy scales for the HF and CASTOR calorimeters have been estimated with an accuracy of 10\%~\cite{Chatrchyan:2011wm} and 15\%~\cite{CMS-DP-2016-006,Sirunyan:2017nsj}, respectively.  The systematic uncertainties are estimated as the change in the fiducial cross section after varying the reconstructed energy in HF and CASTOR by the energy scale uncertainty upwards and downwards.
\item \textbf{Run-to-run variation.} The cross sections are obtained from various runs and the run-to-run variation is taken as an additional source of systematic uncertainty, estimated as the standard deviation of the cross section distribution.
\item \textbf{Integrated luminosity.}  The integrated luminosity of the runs at $B = 3.8\unit{T}$ is determined with an accuracy of 2.3\%~\cite{CMS-PAS-LUM-15-001}.  The integrated luminosity of the $B=0\unit{T}$ runs is rescaled using the ratio of the cross sections for $\xi > 10^{-6}$ at $B=0\unit{T}$ and $B=3.8\unit{T}$.  Because the energy scale and model uncertainties are fully correlated between the $B=0\unit{T}$ and $B=3.8\unit{T}$ samples, the same accuracy of 2.3\% is used to determine the systematic uncertainty due to the integrated luminosity determination at $B=0\unit{T}$.
\end{itemize}

Table~\ref{tab:systematics} gives an overview of the systematic uncertainties.

\begin{table}
\centering
\topcaption{Overview of systematic uncertainties from various sources that contribute to the cross section uncertainty in the two phase space regions considered in this analysis.}
\label{tab:systematics}
\begin{tabular}{lcc}
 & $\sigma(\xi > 10^{-6})$ & $\sigma(\xi_{\mathrm{X}} > 10^{-7} \, \text{or}\, \xi_{\mathrm{Y}} > 10^{-6})$ \\
 & (\unit{mb}) & (\unit{mb}) \\ \hline
Model dependence & 0.68 & 0.39 \\
HF energy scale uncertainty & 0.35 & 0.14 \\
CASTOR energy scale uncertainty & \NA & 0.04 \\
Run-to-run variation & 0.15 &  0.14 \\[\cmsTabSkip]
Total & 0.78 & 0.45 \\[\cmsTabSkip]
Integrated luminosity uncertainty & 1.55 &  1.58 \\
\end{tabular}
\end{table}

\section{Results}

The fully corrected cross sections in phase space domains corresponding to the specific detector acceptances (fiducial cross sections) are given in Table~\ref{tab:corxsec}. The weighted average of the results at $B=3.8\unit{T}$ obtained with the HF calorimeters only is
\begin{equation}
\sigma(\xi > 10^{-6}) = 67.5 \pm 0.8 \syst \pm 1.6\lum\unit{mb},
\end{equation}
with a statistical uncertainty that is smaller than the least significant reported digit.  This can be compared to the cross section  reported by ATLAS for inelastic $\Pp\Pp$ collisions at 13\TeV with $\xi > 10^{-6}$ of $68.1 \pm 0.6 \syst \pm 1.3\lum\unit{mb}$~\cite{Aaboud:2016mmw}.

Averaging the cross sections obtained from runs with the HF and CASTOR calorimeters in the extended phase space yields
\begin{equation}
\sigma(\xi_{\mathrm{X}} > 10^{-7} \, \text{or}\,  \xi_{\mathrm{Y}} > 10^{-6}) = 68.6 \pm  0.5 \syst \pm 1.6\lum\unit{mb},
\end{equation}
also with a statistical uncertainty that is smaller than the least significant reported digit.

\begin{figure}
\centering
\includegraphics[width=0.9\textwidth]{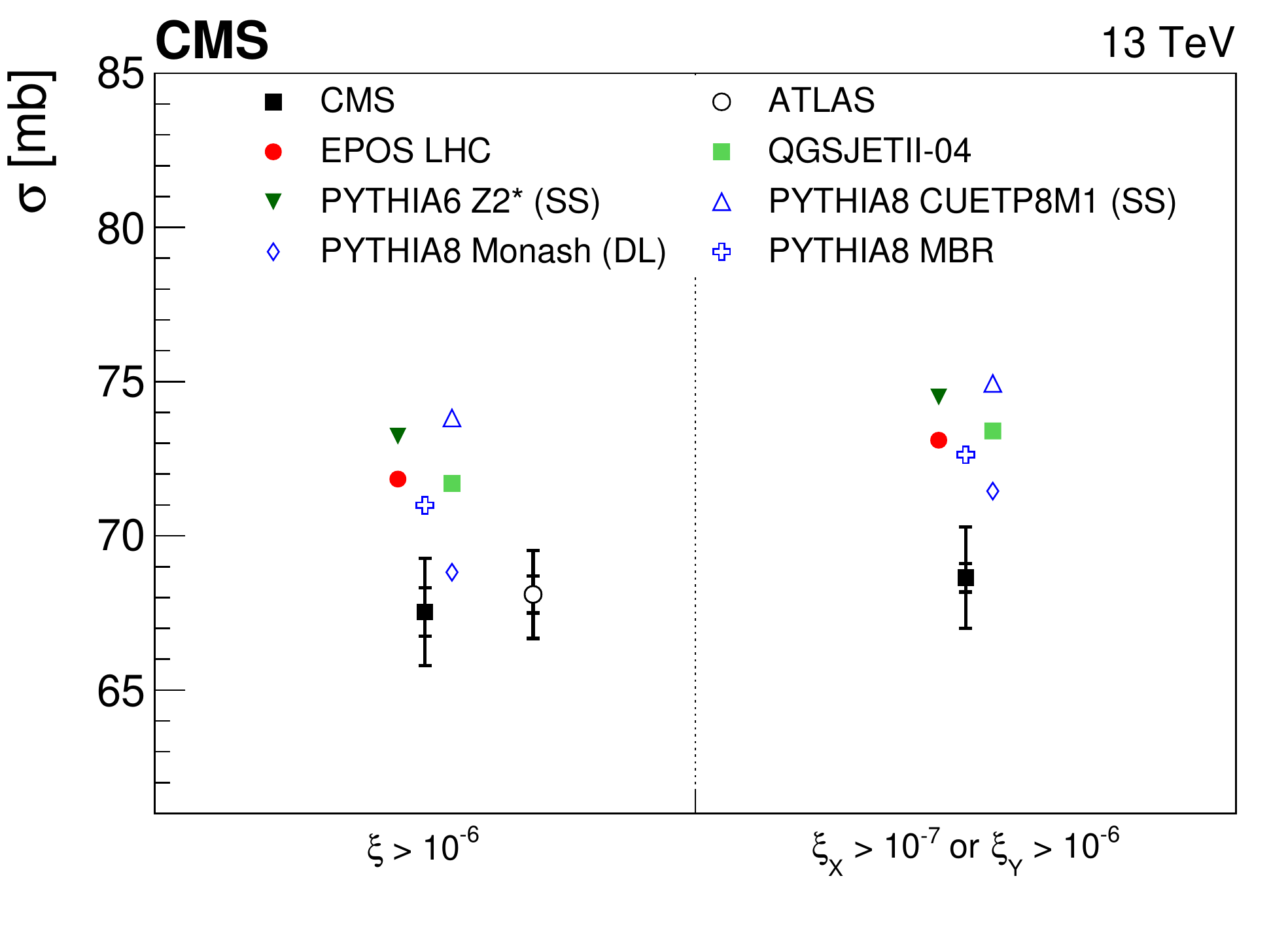}
\caption{Proton-proton inelastic cross section at $\sqrt{s} = 13\TeV$ in two phase space regions, where $\xi = M^2/s$, compared to different models and to the ATLAS result. The inner error bar on the CMS and ATLAS data points indicates the
systematic uncertainty only, while the full error bar reflects the quadratic sum of systematic and integrated luminosity uncertainties. The horizontal placement of the points within each phase space bin is arbitrary.
}
\label{fig:xsec}
\end{figure}

Figure~\ref{fig:xsec} shows the inelastic cross sections in the two phase space domains compared to the predictions of the various models used in this analysis.  Table~\ref{tab:relXS} shows the increase in the cross section from $\xi > 10^{-6}$ to $\xi_{\mathrm{X}} > 10^{-7}$ or $\xi_{\mathrm{Y}} > 10^{-6}$ for the data and the various models. These models are used with their set of tunable parameters as described in Section~\ref{sec:MCmodels}.  Different models and tunes are used to investigate the sensitivity of the fiducial cross section to these parameters, and no uncertainties on individual model predictions are displayed.  The variation between model predictions is mainly due to different descriptions of the diffractive contribution to the cross section.  This can be concluded from the fact that models with the same approach to diffraction predict very similar cross sections (\textsc{Epos} LHC and \textsc{QGSJetII-04}, versus \PYTHIA 6 Z2* (SS) and \PYTHIA 8 CUETP8M1 (SS), respectively).
The relative increase in the inelastic cross section observed in the data is three times larger than the experimental uncertainty. Most models describe reasonably well this small, but significant, relative increase in the inelastic cross section, but in general overpredict their value in both $\xi$ ranges.
The \PYTHIA 8 Monash (DL) model describes data fairly well for $\xi > 10^{-6}$, but predicts a too steep increase for $\xi_{\mathrm{X}} > 10^{-7}$ or $\xi_{\mathrm{Y}} > 10^{-6}$.  Moreover, a comparison of the same models to fiducial and total inelastic cross section measurements at $\sqrt{s} = 7$ and 13\TeV \cite{Abelev:2012sea,Aad:2011eu,Chatrchyan:2012nj,Aad:2014dca,Antchev:2011vs,Antchev:2013gaa,Antchev:2013iaa,Aaboud:2016mmw,Antchev:2017dia} indicates that, while the total inelastic cross section is reasonably well described, the cross section for diffractive dissociation events escaping detection, at masses even lower than
those considered here, is substantially underestimated by the models, leading to an overestimation of the fiducial cross sections.  One may therefore conclude that a model-based extrapolation of the fiducial cross section to the total inelastic phase space would yield a value that is too low for most models.

\begin{table}
\centering
\topcaption{Relative increase in the cross section from the $\xi > 10^{-6}$ to the $\xi_{\mathrm{X}} > 10^{-7}$ or $\xi_{\mathrm{Y}} > 10^{-6}$ fiducial region. The uncertainty in the cross section ratio in data is obtained using systematic uncertainties only, varying each uncertainty source simultaneously for both phase space regions and obtaining the total uncertainty from the variation of the ratio, thus assuming full correlation between both measurements.}
\label{tab:relXS}
\begin{tabular}{lc}
 & Relative cross section increase in \% \\ \hline
Data & 1.64 $\pm$ 0.53 \\
\textsc{Epos} LHC & 1.76 \\
\textsc{QGSJetII-04} & 2.36  \\
\PYTHIA 6 Z2* (SS) & 1.74 \\
\PYTHIA 8 CUETP8M1 (SS) & 1.52  \\
\PYTHIA 8 Monash (DL) & 3.83 \\
\PYTHIA 8 MBR & 2.32 \\
\end{tabular}
\end{table}

\section{Summary}

A measurement of the inelastic proton-proton cross section at $\sqrt{s} = 13\TeV$ with the CMS detector at the LHC has been presented. An inelastic cross section of $67.5 \pm 0.8 \syst \pm 1.6\lum$ mb is obtained for $\xi = M^2/s > 10^{-6}$ (corresponding to $M > 13\GeV$), with $M$ the larger of $M_{\mathrm{X}}$ and $M_{\mathrm{Y}}$, where $M_{\mathrm{X}}$ and $M_{\mathrm{Y}}$ are the masses of the diffractive dissociation systems with negative and positive pseudorapidities, respectively. This result is consistent with a previous measurement in the same phase space \cite{Aaboud:2016mmw}.   In addition, an inelastic cross section of $68.6 \pm 0.5 \syst \pm 1.6\lum\unit{mb}$ is obtained in the enlarged phase space $\xi_{\mathrm{X}} > 10^{-7}$ and/or $\xi_{\mathrm{Y}} > 10^{-6}$ (corresponding to $M_{\mathrm{X}} > 4.1\GeV$ and/or $M_{\mathrm{Y}} > 13\GeV$).
The measured cross sections are smaller than those predicted by the majority of models for hadron-hadron scattering, as previously observed in pp collisions at $\sqrt{s} = 7\TeV$ \cite{Chatrchyan:2012nj}. In contrast, the same models generally describe reasonably well the measurements of the total inelastic cross section at $\sqrt{s} = 13\TeV$  \cite{Aaboud:2016mmw,Antchev:2017dia}.  Given that the difference between the two sets of measurements is dominated by the contribution from low-mass diffractive processes, the data-model discrepancies observed here suggest a theoretical underestimation of the cross section for such events.

\begin{acknowledgments}

We congratulate our colleagues in the CERN accelerator departments for the excellent performance of the LHC and thank the technical and administrative staffs at CERN and at other CMS institutes for their contributions to the success of the CMS effort. In addition, we gratefully acknowledge the computing centers and personnel of the Worldwide LHC Computing Grid for delivering so effectively the computing infrastructure essential to our analyses. Finally, we acknowledge the enduring support for the construction and operation of the LHC and the CMS detector provided by the following funding agencies: BMWFW and FWF (Austria); FNRS and FWO (Belgium); CNPq, CAPES, FAPERJ, and FAPESP (Brazil); MES (Bulgaria); CERN; CAS, MoST, and NSFC (China); COLCIENCIAS (Colombia); MSES and CSF (Croatia); RPF (Cyprus); SENESCYT (Ecuador); MoER, ERC IUT, and ERDF (Estonia); Academy of Finland, MEC, and HIP (Finland); CEA and CNRS/IN2P3 (France); BMBF, DFG, and HGF (Germany); GSRT (Greece); OTKA and NIH (Hungary); DAE and DST (India); IPM (Iran); SFI (Ireland); INFN (Italy); MSIP and NRF (Republic of Korea); LAS (Lithuania); MOE and UM (Malaysia); BUAP, CINVESTAV, CONACYT, LNS, SEP, and UASLP-FAI (Mexico); MBIE (New Zealand); PAEC (Pakistan); MSHE and NSC (Poland); FCT (Portugal); JINR (Dubna); MON, RosAtom, RAS, RFBR and RAEP (Russia); MESTD (Serbia); SEIDI, CPAN, PCTI and FEDER (Spain); Swiss Funding Agencies (Switzerland); MST (Taipei); ThEPCenter, IPST, STAR, and NSTDA (Thailand); TUBITAK and TAEK (Turkey); NASU and SFFR (Ukraine); STFC (United Kingdom); DOE and NSF (USA).

\hyphenation{Rachada-pisek} Individuals have received support from the Marie-Curie program and the European Research Council and Horizon 2020 Grant, contract No. 675440 (European Union); the Leventis Foundation; the A. P. Sloan Foundation; the Alexander von Humboldt Foundation; the Belgian Federal Science Policy Office; the Fonds pour la Formation \`a la Recherche dans l'Industrie et dans l'Agriculture (FRIA-Belgium); the Agentschap voor Innovatie door Wetenschap en Technologie (IWT-Belgium); the F.R.S.-FNRS and FWO (Belgium) under the "Excellence of Science - EOS" - be.h project n. 30820817; the Ministry of Education, Youth and Sports (MEYS) of the Czech Republic; the Council of Science and Industrial Research, India; the HOMING PLUS program of the Foundation for Polish Science, cofinanced from European Union, Regional Development Fund, the Mobility Plus program of the Ministry of Science and Higher Education, the National Science Center (Poland), contracts Harmonia 2014/14/M/ST2/00428, Opus 2014/13/B/ST2/02543, 2014/15/B/ST2/03998, and 2015/19/B/ST2/02861, Sonata-bis 2012/07/E/ST2/01406; the National Priorities Research Program by Qatar National Research Fund; the Programa Severo Ochoa del Principado de Asturias; the Thalis and Aristeia programs cofinanced by EU-ESF and the Greek NSRF; the Rachadapisek Sompot Fund for Postdoctoral Fellowship, Chulalongkorn University and the Chulalongkorn Academic into Its 2nd Century Project Advancement Project (Thailand); the Welch Foundation, contract C-1845; and the Weston Havens Foundation (USA).

\end{acknowledgments}

\bibliography{auto_generated}

\cleardoublepage \appendix\section{The CMS Collaboration \label{app:collab}}\begin{sloppypar}\hyphenpenalty=5000\widowpenalty=500\clubpenalty=5000\vskip\cmsinstskip
\textbf{Yerevan~Physics~Institute,~Yerevan,~Armenia}\\*[0pt]
A.M.~Sirunyan, A.~Tumasyan
\vskip\cmsinstskip
\textbf{Institut~f\"{u}r~Hochenergiephysik,~Wien,~Austria}\\*[0pt]
W.~Adam, F.~Ambrogi, E.~Asilar, T.~Bergauer, J.~Brandstetter, E.~Brondolin, M.~Dragicevic, J.~Er\"{o}, A.~Escalante~Del~Valle, M.~Flechl, M.~Friedl, R.~Fr\"{u}hwirth\cmsAuthorMark{1}, V.M.~Ghete, J.~Grossmann, J.~Hrubec, M.~Jeitler\cmsAuthorMark{1}, A.~K\"{o}nig, N.~Krammer, I.~Kr\"{a}tschmer, D.~Liko, T.~Madlener, I.~Mikulec, E.~Pree, N.~Rad, H.~Rohringer, J.~Schieck\cmsAuthorMark{1}, R.~Sch\"{o}fbeck, M.~Spanring, D.~Spitzbart, A.~Taurok, W.~Waltenberger, J.~Wittmann, C.-E.~Wulz\cmsAuthorMark{1}, M.~Zarucki
% ...
\vskip\cmsinstskip
% ...
1:~Also at~Vienna~University~of~Technology,~Vienna,~Austria\\
% ...
\end{sloppypar}
\end{document}